# *Unsupervised data driven approaches to Raman imaging through a multimode optical fiber*

***LIAM COLLARD[1] , MOHAMMADRAHIM KAZEMZADEH[1] , MASSIMO DE VITTORIO[1,2,*], FERRUCCIO PISANELLO[1 *]***

***1Istituto Italiano di Tecnologia, Center for Biomolecular Nanotechnologies, Arnesano, LE 73010, Italy***

***2Dipartimento di Ingegneria Dell'Innovazione, Università del Salento, Lecce 73100, Italy***

****These authors jointly supervised and are co-last authors of this work***

***liam.collard@iit.it ferruccio.pisanello@iit.it***

**Abstract**

Raman spectroscopy is a label-free, chemically specific optical technique which provides detailed information about the chemical composition and structure of the excited analyte. Because of this, there is growing research interest in miniaturizing Raman probes to reach deep regions of the body. Typically, such probes utilize multiple optical fibers to act as separate excitation/collection channels with optical filters attached to the distal facet to separate the collected signal from the background optical signal from the probe itself. Although these probes have achieved impressive diagnostic performance, their use is limited by the overall size of the probe, which is typically several hundred micrometers to millimeters. Here, we show how a wavefront shaping technique can be used to measure Raman images through a single, hair thin multimode fiber. The wavefront shaping technique transforms the tip of the fiber to a sub-cellular spatial resolution Raman microscope. The resultant Raman images were analyzed with a variety of state-of-the-art statistical techniques including PCA, t-SNE, UMAP and k-means clustering. Our data-driven approach enables us to create high quality Raman images of microclusters of pharmaceuticals through a standard silica multimode optical fiber.

**Introduction**

Raman spectroscopy is a powerful technique used across the life sciences to measure the chemical composition of a sample [1]. When a photon interacts with a molecule, chemical bonds may be excited to a higher vibrational state and the resultant scattered photon is shifted in wavelength. The result is a non-destructive, label-free, chemically specific spectroscopic technique able to assess the molecular composition of a sample with high sensitivity. Raman microscopes are now widely commercially available and are used for a diverse range of applications, including confocal Raman imaging, single cell Raman spectroscopy and chemical sensing. However, there is also substantial interest to miniaturize Raman microscopes for surgical applications, where large bulky microscopic systems are impractical [2].

In this regard, several probes based on multimode optical fibers have been developed, aiming at overcoming two distinct challenges: (i) the large background signal from the waveguide itself which is typically between 1 and 2 orders of magnitude larger than that of the analyte; (ii) engineering the photonic interaction between output light from the fiber which is spatially-scrambled at the output of the fiber. To circumvent these issues, researchers have developed probes based on fiber bundles with optical filters and lenses attached to the distal facet of the waveguide [3–5]. This allows both the background to be eliminated and for spatial resolution to be recovered; however, this increases the overall footprint of the probe, typically to the millimeter scale.

On the other hand, by applying a wavefront shaping technique [6], a single multimode optical fiber can be transformed to a sub-cellular resolution, raster scanning endoscope [7–15], without the requirement for any optical elements attached to distal facet. Raman imaging of microbeads, cells and micrometer scale clusters of pharmaceuticals through a multimode optical fiber has been demonstrated. However these studies typically use visible excitation wavelengths and measure only the high wavenumber Raman shift region 2800 $cm^{-1}$ -3100 $cm^{-1}$, where the silica background is comparatively low [16–18]. For surgical applications, e.g. cancer diagnosis, it may be advantageous to work in the fingerprint region as this provides greater specificity in terms of diagnostic performance. A notable study performed Raman imaging in the fingerprint region through a sapphire multimode fiber [19] which exhibits a background signal with sharp, clearly defined peaks. However a sapphire fiber is not as flexible as silica and therefore may not be well suited to surgical environments.

Here, we demonstrate that state-of-the-art statistical approaches can separate the silica background from the analyte signal and perform Raman imaging in the fingerprint region. Using a bespoke Raman wavefront engineering setup based on a phase-only spatial light modulator (SLM) and 785 nm excitation we raster scanned a diffraction-limited focus through a multimode fiber (MMF) on micrometer scale clusters of paracetamol and ibuprofen. The same MMF was used to collect scattered photons and reconstruct hyperspectral images, extracted at high spatial resolution through the application of a set of different statistical approaches, including polynomial baseline subtraction, principal component analysis, k-means clustering, t-distributed stochastic neighbor embedding (t-SNE) and uniform manifold approximation and projection (UMAP).

## Methods

### Optical train

The experimental setup used to measure Raman images through a multimode optical fiber is shown in **Figure 1**. A 785 nm laser (Toptica, IBeam smart) passed through a laser line filter and had its polarization rotated by a half wave plate to match the requirement of the SLM. The beam was then expanded by a telescope comprised of lenses L1 and L2. The reflected light by the SLM passed through a periscope into the microscope body (CERNA) and a de-magnifying telescope (lenses L3 and L4) reduced the beam size to slightly underfill the back aperture of a microscope objective MO1 (numerical aperture NA=0.3). This effectively reduced the NA of the objective so that it better matched with the Raman imaging MMF (NA=0.22), therefore optimizing the power efficiency of the system. The transmission of the MMF was collected by objective MO2 and focused onto a charge-coupled device (CCD) by lens L6. The position of MO2 could be adjusted in the z axis to set the imaging plane of the MMF. Before performing the effective Raman imaging, the transmission matrix (TM) of the MMF was measured, with a quartz slide placed on the sample holder (no pharmaceutical deposited). After TM measurement, a 50:50% paracetamol: ibuprofen powder was prepared by a pharmaceutical mortar and deposited on the quartz slide. Raman scattering from the sample plane was then collected by the same MMF while Raster scanning a focused spot generated with the holographic algorithm described in next paragraph. The resulting signal was transmitted a long-pass dichroic and passed through a notch filter before being focused onto a patch fiber to couple the Raman signal into a spectrometer (IHR320 (Horiba) equipped with a SYNAPSE CCD cooled at -60 ºC,  slit set at 200 µm, grating 600 lines/mm). An additional CCD monitored the reflection of the proximal facet of the fiber through the 90:10 beam splitter for alignment purposes, while light illumination was used to image the sample for alignment purposes.

### Wavefront correction algorithm

To control the propagation of light through the multimode fiber, a wavefront shaping technique was used, implemented with custom MATLAB software. We measured both the amplitude and phase of the output speckle pattern and then modulated both using the SLM. To determine the amplitude modulation, a set of complex numbers $Z^{amp}_{x_{in},y_{in}}$ were defined with the amplitude indexed by $x_{in}$ and $y_{in}$

$$Z^{amp}_{x_{in},y_{in}} = A_{x_{in},y_{in}}(x_{in}, y_{in})$$

Where $A_{x_{in},y_{in}} = 0$ except at $(x_{in}, y_{in})$ i.e. $A_{x_{in},y_{in}}(x_{in}, y_{in}) = 1$, with $x_{in}$ and $y_{in}$ being the Cartesian coordinates of the input facet. Each complex number $Z^{amp}_{x_{in},y_{in}}$ was then inverse Fourier transformed and had its angle taken to give a blazed grating phase mask $\phi^{amp}_{x_{in},y_{in}}$, which, when applied to the SLM, generates a focused spot ($B^{amp}_{x_{in},y_{in}}$) scanning the different input positions $(x_{in}, y_{in})$ on the proximal fiber facet. The output amplitude scrambled by propagation in the MMF was then measured at every output point $(x_{out}, y_{out})$ in the imaging field of view for every input pair position $(x_{in}, y_{in})$ and stored as a matrix $A_{x_{out},y_{out}}$.

For the phase measurement, another set of phase masks was generated where an input point was selected as an internal phase reference beam, i.e. generating a first focused spot scanning the $(x_{in}, y_{in})$ plane plus a second focused spot in a fixed position $\left(x_{ref}, y_{ref}\right)$ [20] operating as a phase reference. In this case, a set of complex numbers $Z_{x_{in},y_{in},p}$ were defined with the amplitude indexed by $x_{in}$ and $y_{in}$ with phase shifted in four steps $p$=0, $\frac{\pi}{2}$,$\pi$, $\frac{3\pi}{2}$,:

$$Z^{phase}_{x_{in},y_{in},p} = A_{x_{in},y_{in}}(x_{in}, y_{in}) * e^{iP_{x_{in},y_{in},p}}$$

Therefore $A_{x_{in},y_{in}} = 0$ except at $(x_{in}, y_{in})$ and $\left(x_{ref}, y_{ref}\right)$ i.e. $A_{x_{in},y_{in}}(x_{in}, y_{in}) = 1$, and $A_{x_{in},y_{in}}\left(x_{ref}, y_{ref}\right) = 1$ and $P_{x_{in},y_{in},p} = 0$ except at $(x_{in}, y_{in})$ where $P_{x_{in},y_{in},p}(x_{in}, y_{in}, p) = p$. Again, this was inverse Fourier transformed to give a blazed grating phase mask $\phi^{phas}_{x_{in},y_{in},p}$ which when applied to the SLM generated a focused spot scanning the proximal fiber facet $B^{phase}_{x_{in},y_{in},p}$ and fixed reference spot $B^{ref}_{x_{ref},y_{ref}}$. $\phi^{phase}_{x_{in},y_{in},p}$ was then sequentially applied to the SLM and then for every input and output pair $(x_{in}, y_{in})$ and $(x_{out}, y_{out})$ the phase shift $p_{opt_{x_{out},y_{out}}}$ giving the maximum intensity at $(x_{out}, y_{out})$ was stored. Finally to focus the light at $(x_{out}, y_{out})$ a phase mask $\phi^{focus}_{x_{out},y_{out}}$ was defined, as the angle of the inverse Fourier transform of the complex matrix $Z_{opt} = A_{x_{out},y_{out}} * e^{ip_{opt_{x_{out},y_{out}}}}$.

We found that this approach typically achieved a focusing efficiency between 40 and 50 percent which is typical of comparable wavefront engineering/MMF setups [16]. Examples of $\phi^{phase}_{x_{in},y_{in},p}$ and $\phi^{focus}_{x_{out},y_{out}}$ are shown in the insets of **Figure 1** alongside the corresponding speckle patterns.

The relationship between Raman signal and focusing efficiency is as follows. If the field of view (FOV) of the fiber is covered with two materials with reference spectra $S_A$ and $S_B$ with coverage rates $C_A$ and $C_B$ The relationship between resultant Raman signal (excluding silica background) and focusing efficiency $\lambda$ can be described by the following equation

$$Signal = \lambda S_A + (1 - \lambda)\ C_A S_A + (1 - \lambda)\ C_B S_B \qquad (1)$$

When the focus spot does not excite any material, the signal is given by

$$Signal = (1 - \lambda)\ C_A S_A + (1 - \lambda)\ C_B S_B \qquad (2)$$

**Spectral pre-processing**

Raman spectra were processed in MATLAB using a variety of built in functions and in-house software. Spectra were initially recorded between wave numbers 600 $cm^{-1}$ and 2400 $cm^{-1}$. Firstly, the spectral region of interest (ROI) was cropped to only include wave numbers 1100 $cm^{-1}$ to 1800 $cm^{-1}$ as outside of this region the the signal is either masked by large silica peaks at 800 $cm^{-1}$ or 1050 $cm^{-1}$ or silent for the here-employed chemicals. After this, cosmic ray spikes were removed using median filtering. The spectra were then normalized to the silica peak at 1185 $cm^{-1}$ to account for small deviations in excitation power (related to the wavefront correction algorithm). At this point, the next steps in the pre-processing were dependent on the analysis methods and are outlined below.

## Results

In this article, we aim to develop and apply state-of-the-art statistical approaches for assembling Raman images of microparticles through a multimode optical fiber. Our prospective analytical pipeline is illustrated in **figure 1B**. i) SLM activated raster scanning of fiber FOV to measure Raman image. ii) Spectral pre-processing. iii) Unsupervised interpretable (PCA) and non-linear data projection. iv) Qualitative evaluation and microparticle classification. By correlating spatial features found in the non-linear analysis with those found by PCA we are able to optimize both sensitivity and interpretability of our method.

### Raman imaging based on peak intensity through MMFs

For benchmarking purposes, we firstly make Raman images by measuring the intensity of multiple peaks of interest of ibuprofen and paracetamol, which is a typical straightforward method to analyze Raman images. This technique requires a prior knowledge of the analyte spectra and is thus supervised. For paracetamol, we use 1375 $cm^{-1}$, 1565 $cm^{-1}$ and 1649 $cm^{-1}$ and for ibuprofen the 1339 $cm^{-1}$ 1457 $cm^{-1}$ and 1609 $cm^{-1}$ lines as markers for chemical content. These peaks were chosen as in this region silica background is lower. Prior to this, a baseline was removed using alternating least squares fitting on a reduced region of spectral interest (1300 $cm^{-1}$ to 1750 $cm^{-1}$). The complete process is shown step by step in **Figure 2A.** Peak intensity was then measured by numerically integrating each peak. The peak positions were confirmed by measuring reference spectra (**Figure 2B-C**) through MO1.

To assess the quality of the Raman imaging, we compared images generated from Raman spectra with the bright field images shown in **Figure 2D**. The bright field images were measured by transmitting white light through the fiber and monitoring the transmission on CCD2. **Figure S1** shows an annotated version of the brightfield images where each chemical cluster has been labelled. In **Figure 2E-2F** we attempted to make the Raman image of paracetamol and ibuprofen by summing the peaks of interest (that is $I_{1375} + I_{1565} + I_{1649}$ and $I_{1339} + I_{1457} + I_{1609}$). This results in some imaging capacity as features from the transmissive images are also visible in the Raman images, however some clusters are not visible in the Raman images (e.g. the large cluster on top right of image 1 (cluster 1.1 identified in **Figure S1**). A further issue is that this technique is insensitive to overlaying peaks (e.g. the ibuprofen peak at 1609 $cm^{-1}$ and paracetamol peak at 1616 $cm^{-1}$) and thus it is challenging to image each chemical species. Overall, the sensitivity of this technique is quite weak with features in transmissive images not visible in the Raman images.

To try to circumvent this, we took a ratiometric approach and use $\frac{I_{1375} + I_{1565} + I_{1649}}{I_{1339} + I_{1457} + I_{1609}}$, $\frac{I_{1375}}{I_{1339}}$, $\frac{I_{1565}}{I_{1457}}$, $\frac{I_{1649}}{I_{1609}}$, and $\frac{I_{1375} * I_{1565} * I_{1649}}{I_{1339} * I_{1457} * I_{1609}}$ (**Figure 2G-2K**), as markers for paracetamol/ibuprofen. The color scale is set so that red regions correspond to paracetamol and blue to ibuprofen. The resultant image quality is highly dependent on the choice of ratiometric, for instance $\frac{I_{1565}}{I_{1457}}$ gives very clear images of ibuprofren whereas $\frac{I_{1375}}{I_{1339}}$ better resolves the paracetamol. Alternatively, ratiometrics based on combination of peaks $\frac{I_{1375} + I_{1565} + I_{1649}}{I_{1339} + I_{1457} + I_{1609}}$ and $\frac{I_{1375} * I_{1565} * I_{1649}}{I_{1339} * I_{1457} * I_{1609}}$ appear to have better sensitivity to both chemical elements and overall also results in images with better correspondence to the transmissive images, likely due to the overall higher amount of Raman signal analyzed. Nevertheless, several features in the transmissive image are not resolved and there are also uncertainties in the images. For example, in image two, no single ratiometric captures contrast between ibuprofen and paracetamol.

It should be noted that all spectra contain contributions from both ibuprofen and paracetamol (following equation 1). As the unfocussed light is randomly distributed over the field of view for each wavefront, small fluctuations in this signal are also evident in the spectra. In turn, these fluctuations are also dependent on the amount of material within the field of view. To remedy this, one may try to increase the focusing efficiency of the system. Our focusing efficiency value is typical for SLM based wavefront shaping systems without polarization control and complex modulation. Polarization control and complex modulation can produce a near perfect focus at the fiber output [21] at the cost of addition complexity on the system and a less power efficient optical train. We have also aimed to keep the exposure time per spectra low (two seconds) so that the Raman images can be acquired in a reasonable timescale. In principle by increasing the exposure time, the contrast may be improved as the signal-to-noise ratio increases.

**PCA based imaging**

To increase the contrast of the Raman images we look to gather information from bands at lower wavenumbers, effectively increasing the overall optical signal from which the images are made. Unfortunately, up to approximately 1400 $cm^{-1}$ the silica background from the fiber is at least one order of magnitude higher than that of the pharmaceuticals. The silica background also varies across different acquisitions, since the reflected 785 nm light from the material excites additional background signal. At these wavenumbers it is therefore, very difficult to correct the silica background either by experimental or mathematical curve fitting. Our proposed solution is to apply multivariate analysis to decouple the analyte signal from the silica. By associating specific principal components to variations in either analyte/silica variations we aim to be able to probe lower wavenumbers and thus increase the utilized Raman signal. As an added advantage, PCA does not require a priori knowledge of the sample at the distal side of the fiber and is able to extract the pure component spectra.

Our preprocessing pipeline follows that in **Figure 2** up to the baseline removal. After normalization, the mean spectra was subtracted from the dataset and we then applied principal component analysis (MATLAB). The advantage of the PCA technique is evident from the coefficients and scores of the three Raman images shown in **Figure 3**. For example, in image 1, principal component 2 is almost identical to the reference spectra for paracetamol, similarly principal component 5 represents all ibuprofen peaks after and including 1184 $cm^{-1}$. The other principal components shown also represent pharmaceutical peaks. PC1 and PC3 appear to also be representative of paracetamol and deviations in silica background (slope between 1100 $cm^{-1}$ and 1300 $cm^{-1}$). PC4 is a difference between between paracetamol and ibuprofen with a silica background slope. PC6 also appears to represent ibuprofen however is perturbed by silica background. PC7 contains no spatial information and is largely influenced by a spike that was not fully removed.

The PCA approach also appears to be effective at increasing contrast and classifying the microclusters in the second image. PC3 has peaks of all Raman lines associated to ibuprofen from 1212 $cm^{-1}$ to 1467 $cm^{-1}$ whilst PC7 is a difference spectrum between paracetamol and ibuprofen. Alternatively, PC1 represents ibuprofen and a silica variation, whilst PC2 represents a different spectra between ibuprofen and a silica variation. PC4 is a difference spectra covering the approximate spectral region 1100 $cm^{-1}$ to 1400 $cm^{-1}$. PC5 most prominently represents silica. PC6 is also a fairly good representation of paracetamol however is perturbed by silica and, unlike PC7, it does not show bands at 1240 $cm^{-1}$ and 1282 $cm^{-1}$.

As a result of the lack of material within the field of view of the fiber, the third image is somewhat harder to analyze. Nevertheless, the PCA approach has captured clear chemical contrast. PC4 is an extremely good marker for paracetamol, with all bands between 1100 $cm^{-1}$ and 1700 $cm^{-1}$ present. Tentatively, PC2 could be claimed as a marker for ibuprofen minus paracetamol due to the peak at 1339 $cm^{-1}$ however, suffers from a large silica background, making a concrete interpretation of the data more challenging. Again PC1 represents paracetamol and silica. PC3 is relatively similar to PC4 however is perturbed by silica. PC5 is also a good representation of paracetamol. PC6 and PC7 appear to represent non-linear features in the silica background signal.

Due to a complexity of the problem, it is somewhat challenging to put a concrete interpretation on the PCA analysis, however the improvement over the ratiometric/curve fitting approach is clear. When a large amount of material is within the field of view, the PCA approach performs well due to the overall stronger analyte signal across the dataset. In this case, near perfect spectra of analyte may be reconstructed as a high order PCA component. In all images, PC1 appears to represent the dominant chemical species as well as a strong silica background, the edge of the field of view is also represented positively (albeit it weaker than the dominant analyte). This is likely because in this case when the targeted point for focusing is outside of the FOV the light is effectively randomly distributed over the FOV and thus gains a higher analyte signal than when the holographic spot is not exciting any pharmaceutical material. As well as this, the amplitude modulation we perform also weakens the overall excitation power when the light is randomly spread over the FOV, which consequentially introduces a deviation in silica signal. In the first image, PC2 and PC5 are almost perfect representations of the reference spectra with almost all peaks in the spectral region of interest visible and a strong ratiometric correspondence. On the other hand, no principal component for image two could capture the prominent ibuprofen band at 1609 $cm^{-1}$ which is somewhat surprising considering the low silica background in this region. It could be claimed that this peak is most likely represented by a difference on the overlaid paracetamol peak at 1616 $cm^{-1}$. This is a supported by the ratiometric difference between the bands at 1616 $cm^{-1}$ and the neighboring bands at 1659 $cm^{-1}$ and 1565 $cm^{-1}$ when comparing image 2, PC7 and the reference spectra for paracetamol. It is also noteworthy that several PC components indicate simultaneous variation in both silica and analyte signals. These variations are primarily accounted for by two mechanisms. Firstly, the reflection of the excitation light from the pharmaceutical clusters in turn excites additional background signal. Secondly variations in input intensity due to the wavefront correction module between random speckles and focalized. Despite these limitations, the PCA approach clearly outperforms the ratiometric/curve fitting approach in terms of image quality. As well as this, the ability to correlate the scores to reference spectra means it is relatively straightforward to chemically interpret the image, without the requirement for prior knowledge of the analyte.

**k-means classification of chemicals species through an optical fiber**

k-means clustering enables classification of output spectra into distinct clusters by examining similarity across the dataset. We applied it on the correlation matrix between the spectra and applied a Euclidean metric. We selected 4 clusters so that each chemical component, random speckles, and regions with no material could be represented distinctly, we also found 4 as the elbow point of the dataset. The resultant classification images and correlation matrices (sorted in ascending order by k-means classification) are shown in **Figure 4**. The quality of images appears to be qualitatively similar with the PCA approach with the same features visible. As in the PCA, the smaller features (< 10 um) are challenging to resolve with ambiguity in classification (most evident in microclusters 3.1 and 3.2 in image 3). The correlation matrices afford interpretation of similarity of clusters. EG in image 1, clusters 1 and 4 exhibit higher correlation than with 2 or 3. This is likely due to overlapped peaks in paracetamol/ibuprofen spectra.

**Non-linear dimension reduction based imaging**

Alternatively, one possibility to circumvent the sensitivity issues associated to the PCA is to express the spectral data in a reduced dimensionality by applying a non-linear dimensionality reduction technique. A Raman image of microparticles taken through an optical fiber possesses an inherent locality in the dataset structure, whilst the silica background appears to exhibit a global variation across the dataset (see for instance PC5 in image 2). Ultimately, this results in the dispersion of the analyte signal across multiple principal components also linked to silica variation. As well as this, the PCA approach is strongly affected by outliers. This allows us to hypothesize that the non-linear approach is a more sensitive imaging technique with respect to the above-described PCA and k-means approach as it searches for local features in the dataset.

Firstly, we apply 1 dimensional t-SNE [22] to each of the 3 images and form the image for the corresponding score. The strength of the embedding was also evaluated by plotting the score intensity against sorted spectral ID. The improvement over the PCA approach is evident from all 3 images, indeed, even the third image the cluster on the right and center may be resolved (albeit with distortion). The other 2 images show a clear contrast between different chemical regions, absence of material and random speckles.

The values of score shown in **figure 4A** follow a clear evolutional pattern. Consider image 1, the minimum score corresponds to large blue clusters on bottom right the image (Cluster 1.3). From the PCA, we recognize this cluster as paracetamol. Following this, there is a light blue region outside of the FOV, we anticipate this region (where the speckle pattern is random) to also give a high signal from paracetamol as this is the dominant chemical species and the sample is excited by a random speckle pattern. This is followed by a yellow region which from the PCA we ascribe as ibuprofen, and lastly red points represent regions within the FOV with no material. Broadly speaking, the progression could be described as strong paracetamol to paracetamol/ibuprofen to ibuprofen to no material. The other two images follow a similar overall pattern. As well as t-SNE, we also applied UMAP [23] on the same dataset. The techniques are well understood to be extremely similar, however generally UMAP is accepted to be both more efficient and accurate than t-SNE. Indeed for dataset classification t-SNE clustering results in a slightly lower accuracy than UMAP. Overall, this is also clear from our results and UMAP appears to result in a better resolved Raman image than t-SNE.

As the spectra were recorded with a single calibration of the fiber, another possible approach is to apply the multivariate analysis over the entire dataset (all three Raman images). The results of both t-SNE and UMAP applied on all three images are shown in figure 4B-C. The results are almost identical and the improvement in image quality is clear with respect to PCA and k-means, most prominently for the small clusters (<10 um) in the first and third images. In both cases, paracetamol and ibuprofen are represented by the extrema of the scores. As well as this, the upper cluster in image 1 appears to be well resolved by the UMAP with red, orange and dark blue regions corresponding to paracetamol, a composite, and ibuprofen. This approach also allows us to classify the smaller clusters (1.2, 3.1 and 3.2), as PCA components can be directly associated to scores of the UMAP. Tentatively, it could be claimed that in this case t-SNE outperforms UMAP as the clustering plot indicates stronger clustering at the extrema. Nevertheless, it is clear that both techniques are extremely useful methods for analyzing Raman image data.

## Conclusions

State of the art data science based approaches are driving a revolution in Raman sensing and imaging [24,25]. Applying a wavefront shaping technique to perform diffraction limited Raman microscopy at the tip of the fiber and analyzing the resultant data with such approaches, enables to generate high quality Raman images of chemical microclusters through an optical fiber. Non-linear dimension reduction resulted in highest quality Raman images with small clusters (approaching the diffraction limit) resolvable with chemical specificity, while PCA-based analysis enable associating the single cluster to a specific chemical. The approach is also clearly "data hungry", whereby applying non-linear analysis across multiple Raman images resulted in a better image quality (see for instance Figure 5B-C). Consequentially, it is well suited to endoscopic applications where several cellular images may be recorded as a probe is inserted into the brain. We found that PCA worked best on an image by image basis, and whilst it affords a fairly good imaging platform the interpretability of the approach is extremely useful for chemical classification. We have also investigated a k-means based approach for classification, here the resolution appears roughly equivalent to the PCA approach, with an advantage of direct classification.

The techniques presented in this work may also be applied in other hyperspectral imaging techniques. Non-linear Raman techniques such as coherent anti stokes spectroscopy (CARS) offer an opportunity to enhance Raman signal and have demonstrated high frame rate Raman imaging through a multimode fiber [18]. As well as CARS, our results may also be applicable to optical fiber sensors based on other Raman imaging and sensing modalities including surface enhanced Raman spectroscopy (SERS)[26,27], stimulated Raman spectroscopy (SRS)[28] and time gated Raman spectroscopy [29].

Ultimately, we conclude that all three approaches offer valuable information for applied imaging through an optical fiber. Crucially, they demonstrate that a robust statistical approach may overcome the silica background from the fiber and enable imaging in the fingerprint region. We envisage that this study paves the way for new studies incorporating spectral analysis based on cutting edge machine learning approaches, and applications in biological samples.

**Funding**

L.C., M.D.V. and F.P. acknowledge funding from the European Union's Horizon 2020 Research and Innovation Program under Grant Agreement No. 828972. L.C., M.D.V. and F.P. acknowledge funding from the Project “RAISE (Robotics and AI for Socio-economic Empowerment)” code ECS00000035 funded by European Union – NextGenerationEU PNRR MUR - M4C2 – Investimento 1.5 - Avviso “Ecosistemi dell’Innovazione” CUP J33C22001220001. M.D.V. and F.P. acknowledge funding from the European Union's Horizon 2020 Research and Innovation Program under Grant Agreement No 101016787. M.D.V. and F.P. acknowledge funding from the U.S. National Institutes of Health (Grant No. 1UF1NS108177-01). M.D.V and F.P acknowledge funding from European Research Council under the European Union's Horizon 2020 Research and Innovation Program under Grant Agreement No. 966674 M.D.V. acknowledges funding from the European Research Council under the European Union's Horizon 2020 Research and Innovation Program under Grant Agreement No. 692943. M.D.V. acknowledges funding from the U.S. National Institutes of Health (Grant No. U01NS094190). F.P. acknowledges funding from the European Research Council under the European Union's Horizon 2020 Research and Innovation Program under Grant Agreement No. 677683.

**Notes**

M.D.V. and F. Pisanello are founders and hold private equity in Optogenix, a company that develops, produces and sells technologies to deliver light into the brain. MDV: Optogenix srl (I). FP: Optogenix srl (I).

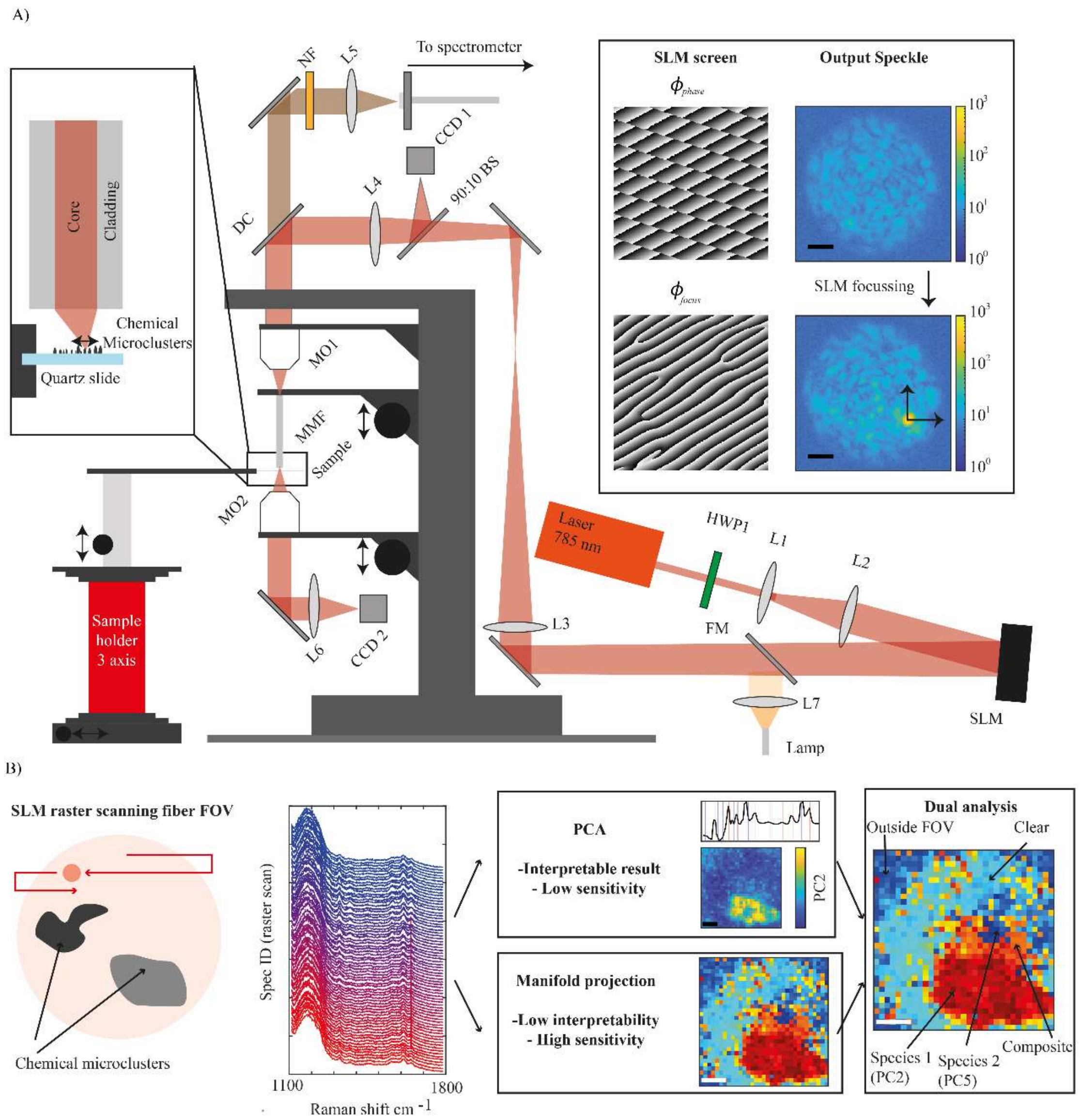

*Figure 1 – A) Optical setup for Raman imaging through a multimode fiber. LL- laser line filter, HWP – half wave plate, L – lens, SLM – spatial light modulator, FM – flip mirror, BS- beam splitter, DC – long pass dichoric mirror, MO – microscope objective, MMF –multimode fiber, CCD – charged coupling device, NF – notch filter. Inset shows typical speckle patterns output from the MMF with and without wavefront shaping focussing (scale bar 10 um). B) Optimized data pipeline for sensitive through fiber Raman imaging.*

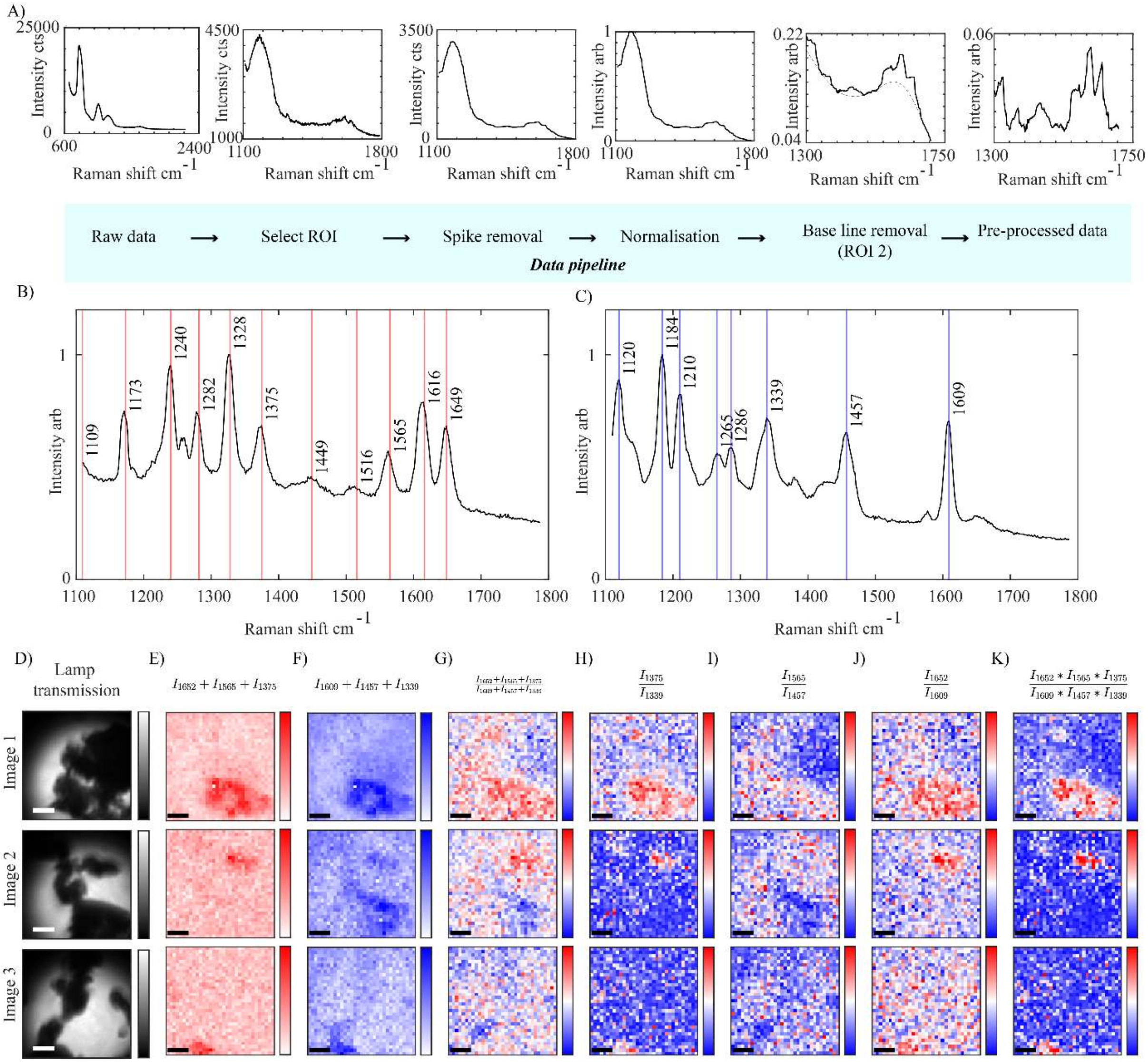

*Figure 2 – Raman imaging via baseline subtraction. A) Pre-processing pipeline. B) Reference spectra for paracetemol taken through MO1 C) Reference spectra for ibuprofen taken through MO1. D) Raman images of paracetamol and ibuprofen clusters based on intensities of various bands of interest.*

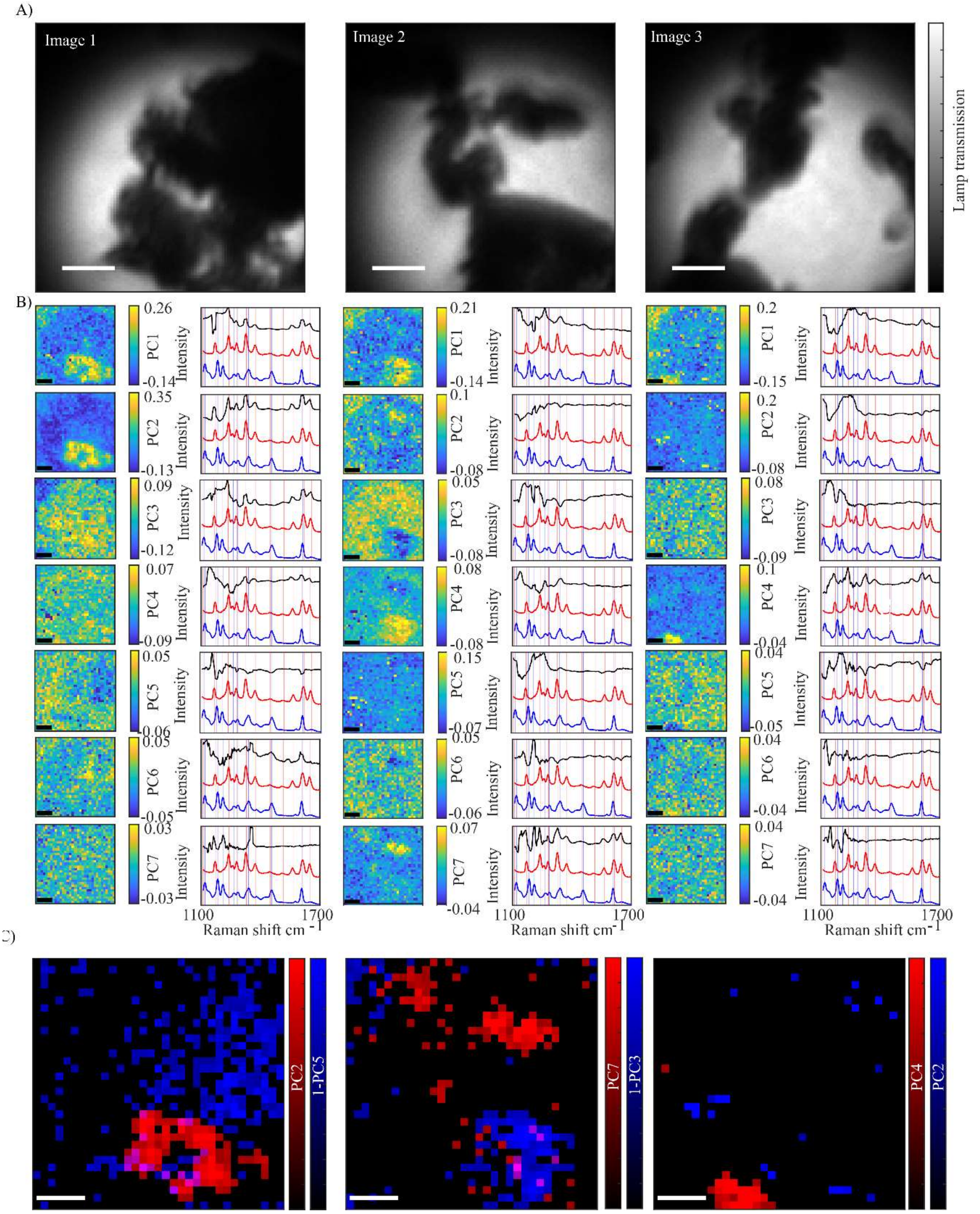

*Figure 3 – Raman imaging via PCA. A) Transmission images with lamp on CCD2. B) Principal component analysis of the Raman image, components 1-7 and the corresponding scores are shown for the 3 images. C) Concatenation of the components identified as paracetamol and ibuprofen.*

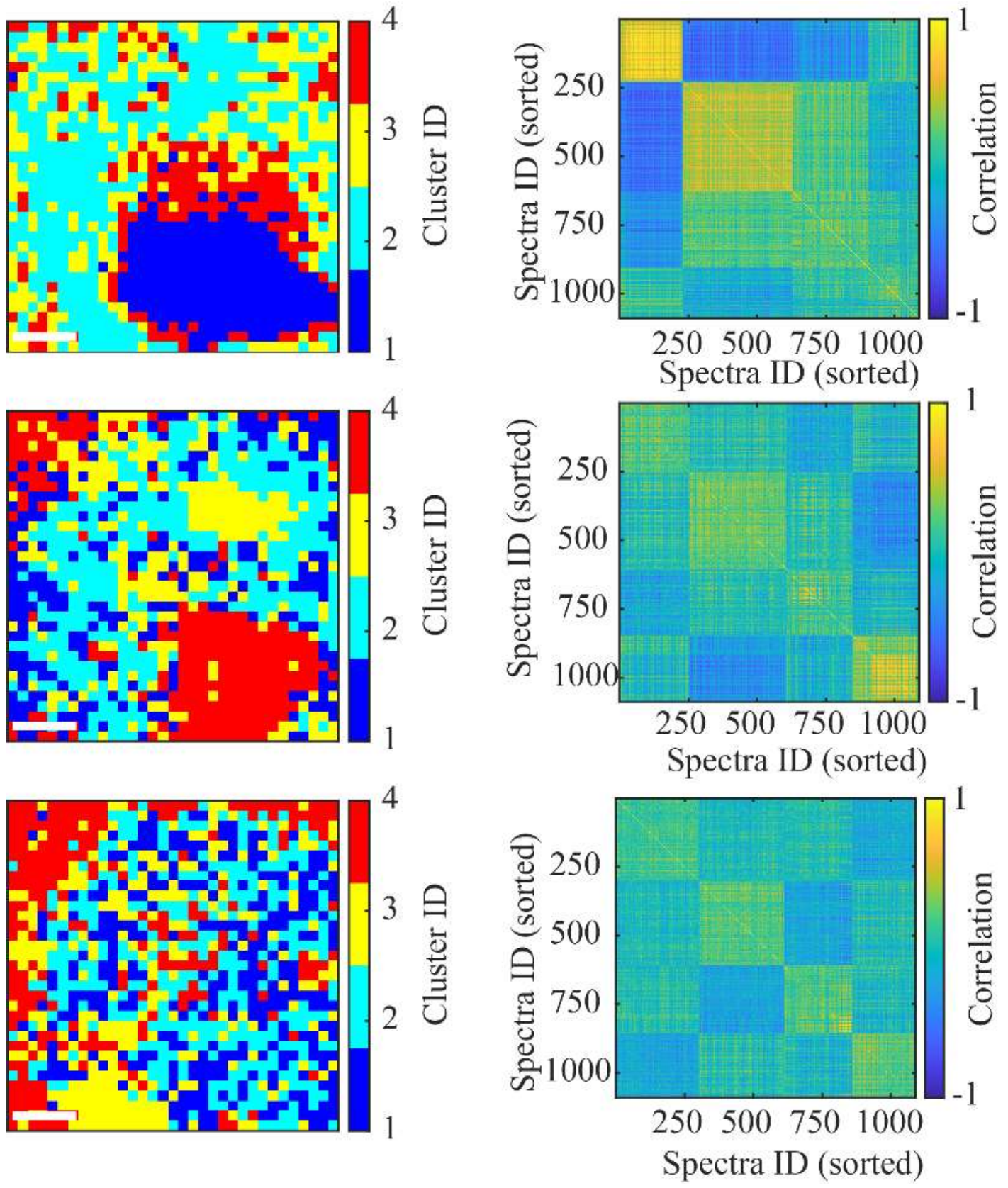

*Figure 4 – Unsupervised classification images based on k-means clustering (correlation matrix) A) Raman images (Image 1-3 (top to bottom)). B) Correlation matrix (sorted by k-means classification).*

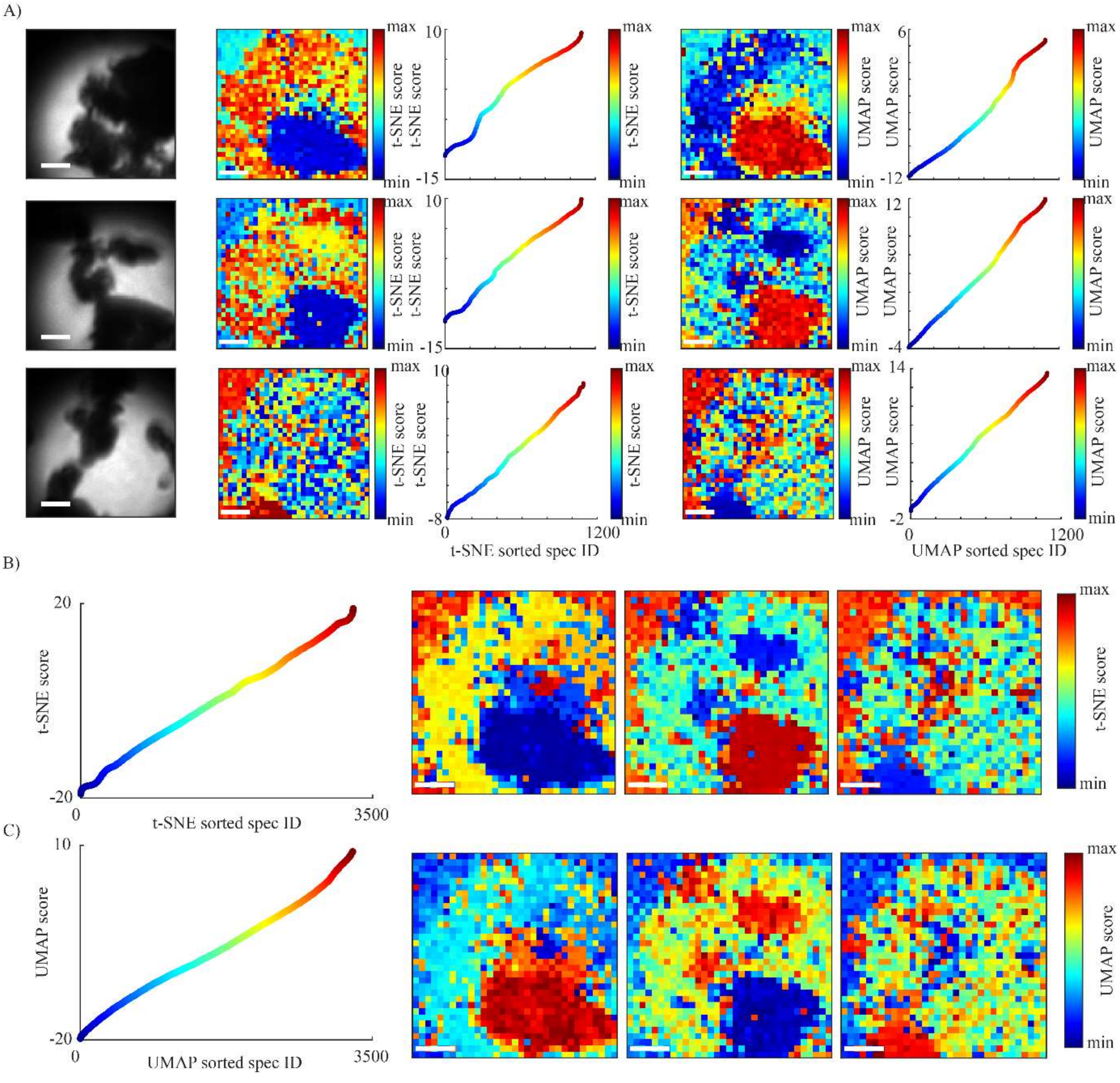

*Figure 5 – Raman imaging via t-SNE and UMAP. A) Raman images through an optical fiber os paracetamol and ibuprofen clusters made with t-SNE and UMAP. B) Raman images made by performing t-SNE on entire dataset (three Raman images) C) Raman images made by performing UMAP on entire dataset (three Raman images).*

# *Supplementary information*

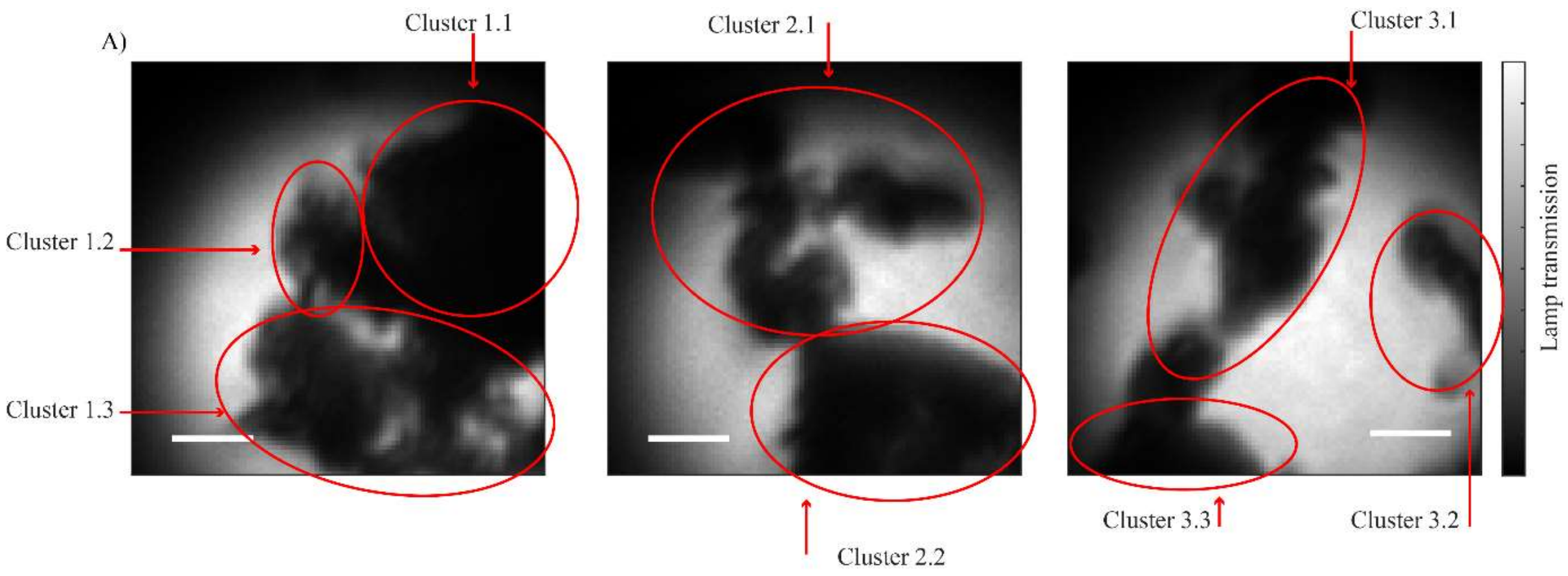

*Figure S1 – Classification of microclusters on bright field images. The combined data analytics shown figures 2 to 5 allow us to classify each cluster as in table S1*

| Cluster ID | Classification | Technique |
|---|---|---|
| 1.1 | Ibuprofen | PCA |
| 1.2 | Composite | PCA+UMAP/t-SNE |
| 1.3 | Paracetamol | PCA |
| 2.1 | Paracetamol | PCA |
| 2.2 | Ibuprofen | PCA |
| 3.1 | Ibuprofen | PCA+UMAP/t-SNE |
| 3.2 | Ibuprofen | PCA+UMAP/t-SNE |
| 3.3 | Paracetamol | PCA |

*Table S1 – Classification of microclusters on bright field images based on the combined data analytics shown figures 2 to 5. The techniques required to classify each cluster are shown in the third column.*